\documentclass{svmult}

\usepackage{graphicx}    

\def\beq{\begin{equation}}
\def\eeq{\end{equation}}

\def\bea{\begin{eqnarray}}
\def\eea{\end{eqnarray}}
\def\ba{\begin{array}}                  
\def\ea{\end{array}}

\newcommand{\muF}{\mu_{F}^{2}}
\newcommand{\muR}{\mu_{R}^{2}}
\newcommand{\muO}{\mu_{0}^{2}}
\begin{document}

\title*{Hard exclusive processes and higher-order QCD corrections}
\author{Kornelija Passek-Kumeri\v{c}ki 
\protect\footnote{Talk given at
the 9th Adriatic Meeting, Dubrovnik 2003.}
        }
\institute{Theoretical Physics Division,
        Rudjer Bo\v{s}kovi\'{c} Institute, \\
        P.O. Box 180, HR-10002 Zagreb, Croatia \\
\texttt{passek@thphys.irb.hr}}
%
%
\maketitle

\begin{abstract}
The short review of the higher order corrections to the
hard exclusive processes is given. 
Different approaches are discussed and the importance of higher-order
calculations is stressed. 
\end{abstract}
\section{Introduction}
\label{sec:intro}

Quantum Chromodynamics (QCD) offers the description of hadrons
in terms of quarks and gluons.
There are two basic ingredients of that picture:
bound state dynamics of hadrons
and
fundamental interactions of quarks and gluons.
While the former is still rather elusive to existing theoretical
tools, the latter is rather well understood. 
The description of the hadronic processes at 
large momentum transfer is realized by making use of
the factorization of high and low energy
(short and long distance) dynamics.
The existence of asymptotic freedom makes then
the high energy part tractable to the perturbative calculation, 
i.e., the perturbative QCD (PQCD).

Exclusive processes are defined as the 
scattering reactions in which the kinematics of all initial
and final state particles are specified,
like, for example, the processes defining
the hadron form factors
($\gamma^* \gamma^{(*)} \to \pi$,
 $\gamma^* \pi \to \pi$,
 $\gamma^* \to \pi \pi$,
  $e p \rightarrow e p$,
     $\cdots$ ),
the two-photon annihilation processes
($\gamma \gamma \to \pi \pi$,
     $\gamma \gamma \rightarrow \overline{p} p$, 
     $\cdots$ ),
the hadron scatterings
($\pi p \rightarrow \pi p$, 
 $p p \rightarrow p p $,
 $\cdots$ ), 
the decays of heavy hadrons 
( $J/\psi \rightarrow \pi \pi \pi$,
  $B \to \pi \pi$, 
       $\cdots$ ) etc.
The hard exclusive reactions, 
i.e., the exclusive reactions at large momentum transfer
(or wide-angle),
can be described by the so-called
hard-scattering picture
\cite{HSP,LepageB80}. 
The basis of this picture is 
the factorization of short and long distance dynamics,
i.e, the factorization of the hard-scattering amplitude
into the elementary hard-scattering amplitude
and hadron distribution amplitudes one for each hadron involved
in the process.
Usually the following standard approximations
are made. Hadron is replaced by the valent Fock state,
collinear approximation, in which hadron constituents
are constrained to be collinear, is adopted,
and the masses are neglected.
For example, in the case of the pion this
leads to replacing the pion by
$\left| \pi \right> 
        \rightarrow 
        \left| q \overline{q} \right> 
        $
(correct flavour structure has to be taken into account),
adopting $ p_{q}= x \: {p}$,
   $p_{\overline{q}}= (1-x) p$, where 
$p$, $p_q$ and $p_{\overline{q}}$ 
are pion, quark and antiquark momenta, respectively, while
$0<x<1$ is the longitudinal momentum fraction,
and taking
$m_q=m_{\overline{q}}=0 \:, m_{\pi}=0$.

Generally, the hard-scattering amplitude is then
represented by the following
convolution formula:
\begin{equation}
{\cal M}(Q^2)
  =  \int_{0}^{1} [dx] \quad
{T_H}(x_{j}, \, {Q^2},\, \muF)
              \quad {\prod_{{h}_{i}}} \,
\Phi_{{h}_{i}}(x_{j}, \, \muF)
\, ,
\end{equation}
\begin{displaymath}
   [dx] = \prod_{j=1}^{{n}_{h_i}} dx_{j} \,
            \delta(1-\sum_{k=1}^{{n}_{h_i}} x_{k})
\, ,
\label{eq:conv}
\end{displaymath}
where $T_H$ is the process-dependent elementary hard-scattering amplitude,
$\Phi_{{h}_{i}}$ is the process-independent
distribution amplitude (DA) of the hadron
${h}_{i}$, $Q^2$ denotes the large momentum transfer
while $\muF$ is the factorization scale at which the
separation between short and long distance dynamics takes
place.

Within this framework 
leading-order (LO) predictions have been obtained for many
exclusive processes.
It is well known, however, that, unlike in QED, the LO predictions in PQCD
do not have much predictive power, and that higher-order corrections
are essential for many reasons.
In general, they have a stabilizing effect reducing the dependence of the
predictions on the schemes and scales. Therefore, to achieve a complete
confrontation between theoretical predictions and experimental data, it is
very important to know the size of radiative corrections to the LO
predictions.

The list of exclusive processes at large momentum transfer analyzed at
next-to-leading order (NLO) is very short and includes only three processes:
the meson electromagnetic form factor
   \cite{FieldGOC81,DittesR81,Sarmadi82,KhalmuradovR85,
       KadantsevaMR86,BraatenT87,MNP99},
the meson transition form factor 
    \cite{AguilaC81,Braaten83,KadantsevaMR86, KrollP02},
and the process
$\gamma \gamma \rightarrow M \overline{M}$ ($M=\pi$, $K$)
\cite{Nizic87}
\footnote{In contrast to the above introduced standard
hard-scattering approach (sHSA), in the so-called modified
hard-scattering approach (mHSA) the Sudakov suppression and the transverse
momenta of the constituents are taken into account.
The LO predictions have again been obtained for number of processes
while at NLO order only the pion transition form factor \cite{MusatovR97}
has been calculated. In order to estimate the NLO correction
in the mHSA, in \cite{StefanisSchK00} the use has been made of the
NLO results for the pion electromagnetic and transition form factors 
obtained using the sHSA.}.

We note here that the meson transition form factor belongs to the
same class of processes as the 
deeply virtual Compton scattering (DVCS) \cite{MullerRGDH94}
($\gamma^* p \rightarrow \gamma^* p$),
which recently has been extensively studied in the context of
general parton distributions (GPDs)
\cite{DVCS}.
Regarding the elementary hard-scattering amplitude, these two
processes, or correspondingly subprocesses
$\gamma^* \gamma^* \rightarrow (q \bar{q})$
and $\gamma^* q \rightarrow \gamma^* q$, differ 
only in kinematic region and are related by crossing. 
Still, the NLO correction to DVCS has been calculated 
independently \cite{MankiewiczPSVW98,JiO98,BelitskyM98}.
Analogous connection exists between the pion electromagnetic
form factor and the deeply virtual electroproduction of mesons (DVEM)
\cite{DVEM}
($\gamma^* p \rightarrow M p$, 
the momentum transfer $t$ between the 
initial and the final proton is negligible, 
while the virtuality of the photon is large).
In the context of subprocesses, there is a connection between 
$\gamma^* (q \bar{q}) \rightarrow (q \bar{q})$ and
$\gamma^* q \rightarrow (q \bar{q}) q$.
In \cite{BelitskyM01} the use has been made of the NLO
results for the pion electromagnetic form factor 
to obtain the NLO prediction for the specific case of DVEM
(electroproduction of the pseudoscalar flavour non-singlet
mesons). 
In this work we mostly discuss the meson form factor calculations.

At the next-to-next-to-leading order (NNLO)
only the $\beta_0$-proportional terms
for the deeply virtual Compton scattering 
and pion transition form factor have been explicitly
calculated \cite{BelitskySch98,MNP01}.
The use of conformal constraints 
made it possible to circumvent the explicit calculation
and to obtain the full NNLO
results for the pion transition form factor \cite{MelicMP02}.

Finally, let us note that apart from the deeply virtual region, 
also the wide angle region has been investigated in the literature 
in the context of the Compton scattering (WACS), as well as, 
electroproduction of mesons (WAEM) \cite{HuangK00}. 
The NLO corrections were calculated only for the WACS 
\cite{HuangKM01,HuangM03}.

In this paper the 
introduction to hard-scattering picture for exclusive
processes is given in Sec. \ref{sec:HSP}.
The characteristic  properties of the PQCD predictions regarding
the importance of higher-order corrections and 
the renormalization scale ambiguities
are explained in Sec. \ref{sec:PQCDprediction}.
Section \ref{sec:higherorder} 
is devoted to the short review of exclusive processes 
calculated to the next-to-leading order (NLO) in the strong coupling constant
$\alpha_S$:
meson electromagnetic form factor 
($\gamma^* M \rightarrow M$),
photon-to-meson transition form factor
 ($\gamma^* \gamma^{(*)} \rightarrow M$),
meson pair production: $\gamma \gamma \rightarrow M \bar{M}$.
In Sec. \ref{sec:NNLO} the 
next-to-next-to-leading order (NNLO) prediction 
for the photon-to-pion transition form factor 
obtained using conformal symmetry constraints is explained.
Finally, in Sec. \ref{sec:concl} the summary and conclusions are given.

\section{Introduction to the hard-scattering picture}
\label{sec:HSP}

Let us explain the basic ingredients of the standard
hard-scattering picture by taking as an example the simplest
exclusive quantity, i.e., 
the photon-to-pion transition form factor $F_{\pi \gamma^{(*)}}$ 
appearing in the amplitude of the process
${\gamma^*} (q_1, \mu) {\gamma^{(*)}} (q_2, \nu) 
\rightarrow \pi (p)$. 
At least one photon virtuality has to be large and we 
take here the simple case : $-q_1^2=Q^2 \gg$ and $q_2^2=0$.
The full amplitude is of the form
\begin{equation}
    \Gamma^{\mu}= i \, e^2 \; F_{\pi\gamma}(Q^{2})
         \; \varepsilon^{\mu \nu \alpha \beta} 
                          \;q_{1\alpha} q_{2\beta}
                          \; \epsilon_{\nu}(q_2)
\, ,
\end{equation}
and the transition form factor
can be represented by a convolution
\begin{equation}
F_{\pi \gamma}(Q^{2})=
     T_{H}( x,Q^{2},\muF)  \,
           \otimes \, 
     \Phi( x, \muF)
\, .
\label{eq:Fpigammaconv}
\end{equation}
Here $A(x) \otimes B(x) \equiv \int_0^1 dx A(x) B(x)$ and 
$\muF$ is a factorization scale.

\begin{figure}
\centerline{\includegraphics[width=9cm]{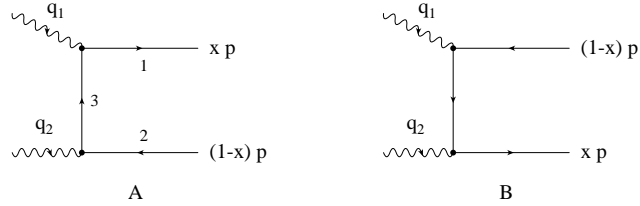}}
 \caption{Lowest-order Feynman diagrams contributing to the
  $\gamma^* \gamma \rightarrow q \overline{q}$ amplitude.}
 \label{f:Tlo}
\end{figure}

\begin{figure}
\centerline{\includegraphics[width=8cm]{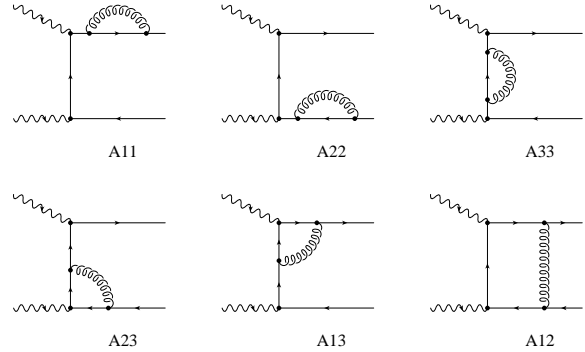}}
 \caption{Distinct one-loop Feynman diagrams contributing to 
  $\gamma^* \gamma \rightarrow q \overline{q}$.}
 \label{f:Tnlo}
\end{figure}

The elementary hard-scattering amplitude
$T_H$ obtained from
$ \gamma^* \, \gamma \rightarrow q \overline{q}$
is calculated using the PQCD.
By definition, $T_H$ is free of collinear singularities and has a
well--defined expansion in $\alpha_S(\muR)$,
with $\muR$ being the renormalization (or coupling constant) scale
of the hard-scattering amplitude.
Thus, one can write
\begin{eqnarray}
T_H(x,Q^2)&=&T_H^{(0)}(x,Q^2) + 
\frac{\alpha_S(\muR)}{4 \pi} T_H^{(1)}(x,Q^2,\muF) 
\nonumber \\ & &
 + \frac{\alpha_S^2(\muR)}{(4 \pi)^2} T_H^{(2)}(x,Q^2,\muF,\muR) +
\cdots
\, .
\end{eqnarray}
The diagrams contributing to LO and representative diagrams
contributing to NLO order are displayed
in Figs. \ref{f:Tlo} and \ref{f:Tnlo}, respectively.
When evaluating the NLO amplitude one encounters
the UV and collinear singularities.
The former are removed by coupling constant ($\alpha_S$) 
renormalization introducing the scale $\muR$, while the latter
are factorized into the DA at the scale $\muF$.

The pion distribution amplitude is
defined in terms of the matrix elements of composite
operators:
$
\langle 0 | \Psi(-z) \gamma^+ \gamma_5 \Omega \Psi(z) | \pi \rangle$.
While the DA form is taken as an 
(nonperturbative) input at some lower scale
$\muO$ ($\Phi(x,\muO)$), 
its evolution to the factorization scale $\muF$ ($\Phi(x,\muF)$)
is governed by PQCD.
The DA can hence be written in a form
\begin{equation}
\Phi(x,\muF)={\phi_V(x,y,\muF,\muO)} \otimes \Phi(y,\muO)
\, ,
\end{equation}
where $\phi_V$ denotes the evolution part of the DA. 
In latter the resummation of $(\alpha_S \ln(\muF/\muO))^n$ terms
is usually included, and $\phi_V$ is obtained by solving
the evolution equation
\begin{equation}
{\muF} \frac{\partial}{\partial {\muF}} \phi_V   =
   V \, \otimes \, \phi_V \: 
\, ,
\label{eq:eveq}
\end{equation}
where 
\begin{equation}
V=\frac{\alpha_S(\muF)}{4 \pi} V_1 + \frac{\alpha_S^2(\muF)}{(4 \pi)} V_2
 + \cdots \:
\label{eq:kernel}
\end{equation}
represents the perturbatively calculable evolution kernel.

One often introduces the distribution amplitude
$\phi$ normalized to unity
$\int_0^1 dx \; \phi(x,\muF) = 1$,
and related to $\Phi(x,\muF)$ by
$\Phi= f_{\pi}/(2 \sqrt{2 N_c}) \, \phi \, $
where $f_{\pi}=0.131$ GeV is the pion decay constant
and $N_c$ is the number of colours.
The solutions of the evolution equation (\ref{eq:eveq}) 
combined with the nonperturbative input can then be written in a form 
of an expansion over Gegenbauer polynomials 
$C_n^{3/2}$ which represent the eigenfunctions of 
the LO evolution equation:
\begin{equation}
\phi(x,\muF)= 6 x (1-x) \left[ 1 + \sum_{n=2}^{\infty}{}' 
     B_n(\muF) \; C_n^{3/2}(2 x -1) \right]
\, .
\end{equation}
Here $\sum{}'$ denotes the sum over even indices.
The nonperturbative input $B_n(\muO)$ as well as
the evolution is now contained in $B_n$ coefficients.
They have a well defined expansion in
$\alpha_S$:
\begin{equation}
B_n(\muF) = B_n^{LO}(\muF) + \frac{\alpha_S(\muF)}{4 \pi} B_n^{NLO}(\muF) 
            + \cdots
\, ,
\end{equation}
where
\begin{eqnarray}
B_n^{LO}(\muF)=f(\muF,\muO, B_n(\muO))
\, ,
& \quad&
B_n^{NLO}(\muF)=g(\muF,\muO, B_{k (k\leq n)}(\muO) )
\qquad
\end{eqnarray}
represent the LO and NLO \cite{Muller94etc} parts 
whose exact form in $\overline{\mbox{MS}}$ factorization scheme
is given in, for example, \cite{MNP99,MNP01}.

As the DA input one often takes the asymptotic function
\begin{equation}
\phi_{as} \equiv \phi(x,\infty)= 6 x (1-x)
\end{equation}
being the solution of the DA evolution equation for
$\muF\to \infty$ and the simplest possibility.
We list here two more choices from the literature
\begin{equation}
  \begin{array}{rl}
   \phi_{CZ} \mbox{\cite{ChernyakZ84}}: &  B_2(0.25 \, \mbox{GeV}^2)=2/3  
            \, ,
            \\[0.1cm]
   \phi_{BMS}\mbox{\cite{BakulevMS01}}: &  B_2(1\,  \mbox{GeV}^2)=0.188  
            \qquad B_4(1\,  \mbox{GeV}^2)=-0.13 
               
\, .
   \end{array}
\end{equation}
The CZ distribution amplitude is nowadays mostly
ruled out (see \cite{MNP99b} and references therein),
and it is believed that even at lower energies
the pion DA is close to asymptotic form but probably
end-point suppressed like the BMS DA.
\begin{figure}
\centerline{\includegraphics[width=8cm]{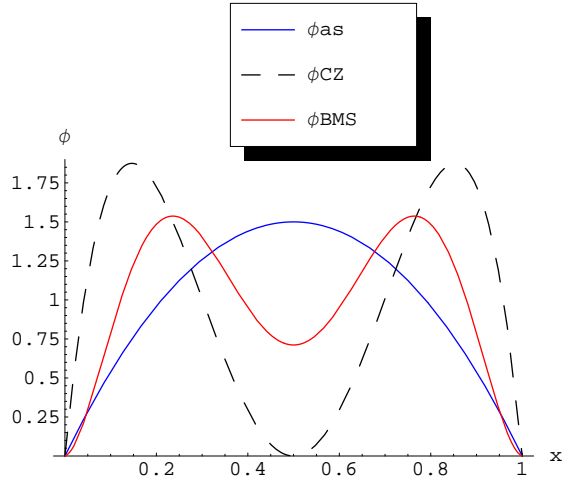}}
\caption{DA candidates.}
\end{figure}

\section{PQCD prediction}
\label{sec:PQCDprediction}

In this section we would like to discuss some properties
inherent to all PQCD predictions.

Let us first briefly discuss the expansion parameter,
i.e., the QCD coupling constant.
The QCD $\beta$ function given by
\begin{equation}
\beta(\alpha_S(\mu^2))=\mu^2 \frac{\partial}{\partial \mu^2} \alpha_S(\mu^2)
=-\frac{\alpha_S^2(\mu^2)}{4 \pi} \beta_0- \cdots
\label{eq:beta}
\end{equation}
is negative
($\beta_0=1/3(11 N_c - 2 n_f)$, $n_f$ is the number of active flavours) 
and theory is asymptotically free.
The usual one-loop solution of the renormalization group equation
(\ref{eq:beta}) is given by
\begin{equation}
\displaystyle \alpha_S(\mu^2)=\frac{4 \pi}{\beta_0 \ln (\mu^2/\Lambda^2)}
\end{equation}
and, obviously,
$\alpha_S(\infty)=0$, 
while,  for example,
$\alpha_S((1)4 \, \mbox{GeV}^2)\approx (0.43)0.3$,
$(\Lambda=0.2 \, \mbox{GeV}, n_f=3)$. 
Obviously, low-energy behaviour of such $\alpha_S$
represents a problem due to the existence of the Landau pole:
$\alpha_S(\Lambda^2)\rightarrow \infty$.
In the literature one can encounter several prescriptions 
to improve $\alpha_S$ in low-energy region.
For example,
``frozen'' coupling constant
\begin{equation}
 \alpha_S(\mu^2)=\frac{4 \pi}{\beta_0 
                  \ln ((\mu^2+m_g^2)/\Lambda^2)}
\end{equation}
or analytical coupling \cite{ShirkovS97}
\begin{equation}
 \alpha_S(\mu^2)=\frac{4 \pi}{\beta_0}\left[ 
                  \frac{1}{\ln (\mu^2/\Lambda^2)}
                 + \frac{\Lambda^2}{\Lambda^2-\mu^2} \right]
\, .
\end{equation}
But even these improved forms of  $\alpha_S$ give
rather large values at lower energies.
Thus, LO QCD predictions do not have much predictive power
and higher-order corrections are important.

Generally, the PQCD amplitude can be written in a form
\begin{equation}
{\cal M}(Q^2) = {\cal M}^{(0)}(Q^2)
+ \frac{\alpha_S(\muR)}{4 \pi} {\cal M}^{(1)}(Q^2)
+ \frac{\alpha_S^2(\muR)}{(4 \pi)^2} {\cal M}^{(2)}(Q^2,\muR)
+ \cdots
\, ,
\label{eq:Mexp}
\end{equation}
where $Q^2$ is some large momentum and as usual $\muR$ represents
the renormalization scale.
The truncation of the perturbative series to finite
order introduces the residual dependence of the results
on the renormalization scale $\mu_R$ and scheme
(to the order we are calculating these dependences can be represented
by one parameter, say, the scale). 
Inclusion of higher order
corrections decreases this dependence.
Nevertheless, we are still left with intrinsic theoretical
uncertainty of the perturbative results.
One can try to estimate this uncertainty
(see, for example, \cite{MNP99}) or
one can try to find the ``optimal'' renormalization scale $\mu_R$ 
(and scheme) on the basis of some physical arguments .
In the latter case, one can 
assess the size of the higher order corrections
and of the expansion parameter. These values can then serve as a sensible
criteria for the convergence of the expansion.

The simplest and widely used choice
for $\mu_R$ is
$\muR = Q^2$,
and 
the justification is mainly pragmatic.
However, physical arguments suggest that the more appropriate
scale $\mu_R$ is lower.
Namely, since each external momentum entering an exclusive reaction
is partitioned among many propagators of the
underlying hard-scattering amplitude, the physical scales
that control these processes are inevitably much softer
than the overall momentum transfer.
There are number of suggestions in the literature.
According to fastest apparent convergence (FAC) procedure
\cite{FAC},
the scale $\mu_R$ is determined
by the requirement that the NLO coefficient in
the perturbative expansion of the physical
quantity in question vanishes, i.e.,
one demands
${\cal M}^{(2)}(Q^2,\muR)=0$.
On the other hand, following the
principle of minimum sensitivity (PMS)
\cite{PMS}
one mimics the independence of the all order expansion
on the scale $\mu_R$, and one chooses the renormalization
scale $\mu_R$ at the stationary point of the truncated
perturbative series:
$d{\cal M}_{\mbox{finite order}}(Q^2,\muR)/d\muR=0$. 
In the Brodsky-Lepage-Mackenzie (BLM)
procedure \cite{BLM83},
all vacuum-po\-la\-ri\-zat\-ion effects from the
QCD $\beta$-function are resummed
into the running coupling constant.
According to BLM procedure,
the renormalization scale best suited
to a particular process in a given order can, in practice, be
determined by setting the scale
demanding that $\beta$-proportional terms should vanish:
\begin{equation}
{\cal M}^{(2)}(Q^2,\muR)= 
\beta_0 \, {\cal M}^{(2,\beta_0)}(Q^2,\muR)
+{\cal M}^{(2,\mbox{rest})}(Q^2)
\end{equation}
and
\begin{equation}
{\cal M}^{(2,\beta_0)}(Q^2,\muR)=0
\, .
\end{equation}

As it is known, the relations between physical observables must
be independent of renormalization scale and scheme
conventions to any fixed order of perturbation theory.
In Ref. \cite{BrodskyL95} was argued that
applying the BLM scale-fixing to perturbative predictions
of two observables in, for example, $\overline{\mbox{MS}}$ scheme
and then algebraically eliminating
$\alpha_{\overline{MS}}$ one can relate
any perturbatively calculable observables without scale and scheme
ambiguity, where the choice of BLM scale ensures
that the resulting ``commensurate scale relation'' (CSR)
is independent of the choice of the intermediate renormalization
scheme.
Following this approach,
in paper by Brodsky {\em et al.}\cite{BrodskyJPR98}
the several exclusive hadronic amplitudes
were analyzed in
$\alpha_V$ scheme,
in which the effective coupling $\alpha_V(\mu^2)$ is defined
from the heavy-quark potential $V(\mu^2)$.
The $\alpha_V$ scheme is a natural, physically based scheme,
which by definition automatically incorporates vacuum polarization
effects.
The $\mu_V^2$ scale which then appears in
the $\alpha_V$ coupling reflects the mean virtuality of the
exchanged gluons.
Furthermore, since $\alpha_V$ is an effective running coupling defined
from the physical observable it must be finite at low momenta,
and the appropriate parameterization of the low-energy region
should in principle be included.
The scale-fixed relation between the
$\alpha_{\overline{MS}}$ and $\alpha_V$
couplings is given by \cite{BrodskyJPR98}
\begin{equation}
\alpha_{\overline{MS}}(\mu_{BLM}^2)
=
\alpha_V(\mu_V^2) \left( 1 + \frac{\alpha_V(\mu_V^2)}{4 \pi}
          \, \frac{8 C_A}{3} + \cdots \right)
\, ,
\end{equation}
where
$\alpha_V(\mu_V^2) $ is 
defined from the heavy-quark potential $V(\mu_V^2)$
 and
\begin{equation}
\mu_V^2 = e^{5/3} \, \mu_{BLM}^2
\, .
\end{equation}

\section{Exclusive processes at higher order: explicit calculations}
\label{sec:higherorder}

As mentioned in Sec. \ref{sec:intro}, only the small number of
exclusive processes have been analyzed in higher orders.
When higher order calculations
are explicitly performed, usually
the dimensional regularization together
with the $\overline{\mbox{MS}}$ renormalization scheme is
applied.

\subsection{
 Photon-to-$\pi$ ($\eta$, $\eta'$) transition form factor}
\label{ssec:tff}

The photon-to-$\pi$ transition form factor appearing
in the amplitude
${\gamma^*} {\gamma} { \rightarrow} {\pi^0 }$
takes the form of an expansion
\begin{eqnarray}
  F_{\pi \gamma}(Q^2)&=&F^{(0)}_{\pi \gamma}(Q^2)
  + \frac{\alpha_S(\muR)}{4 \pi}
  F^{(1)}_{\pi \gamma}(Q^2) 
\nonumber \\ & &
  + \frac{\alpha_S^2(\muR)}{(4 \pi)^2}
  \left[ \beta_0 \, F^{(2, \beta_0)}_{\pi \gamma}(Q^2,\muR) + 
         \cdots \right] 
  + \cdots
\, ,
\end{eqnarray}
where only the parts that can be found in the literature
as explicitly calculated from the contributing Feynman diagrams
are written.

There are 2 LO diagrams contributing to the subprocess
amplitude
$\gamma^* \gamma \rightarrow (q \bar{q})$
and displayed in Fig. \ref{f:Tlo}.
Furthermore, there are 12 one-loop diagrams contributing at NLO
order \cite{AguilaC81,Braaten83,KadantsevaMR86}. 
The representative diagrams are given in Fig. \ref{f:Tnlo}.
In the case of the photon-to-$\eta$ ($\eta'$) transition
form factor  the two-gluon states also contribute
($\gamma^* \gamma \rightarrow (g g)$) giving rise to
6 more diagrams at NLO \cite{KrollP02}.
In \cite{MNP01} the $\beta_0$-proportional NNLO terms
were determined from the 12 two-loop Feynman diagrams obtained
from the one-loop diagrams by adding the gluon vacuum polarization
bubble.

\begin{figure}
\centerline{\includegraphics[width=9cm]{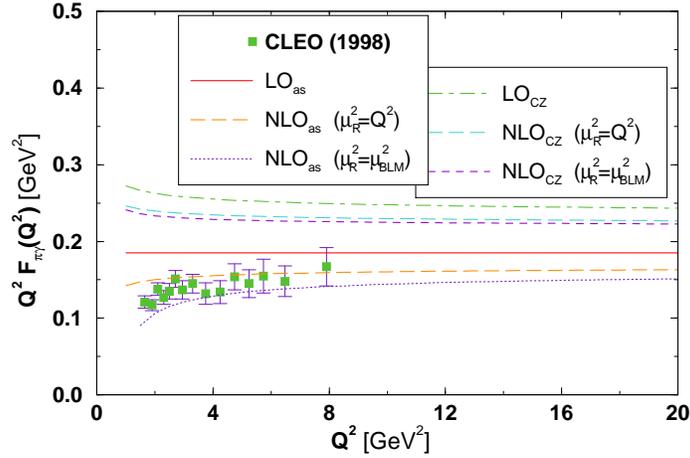}}
\caption{LO and NLO predictions for the photon-to-pion
transition form factor.}
\label{f:num1}
\end{figure}

The numerical predictions for $F_{\pi \gamma}(Q^2)$
are displayed in Fig. \ref{f:num1}.
Obviously, the results obtained using the CZ DA overshoot 
the experimental data.
The BLM scale for the asymptotic DA amounts to
$(\mu_{BLM}^2)^{as} \approx Q^2/9$,
while
$\alpha_S\leq 0.5 \mbox{ for } Q^2>4 \mbox{ GeV}^2$. 
In the $\alpha_V$ scheme for the coupling constant
scale one obtains $(\mu_{V}^2)^{as} \approx Q^2/2$.

\subsection{
 Pion electromagnetic form factor}
\label{ssec:pff}

The spacelike%
\footnote{For the discussion of the timelike form factor 
see, for example,  \cite{BakulevRS00}.}
pion electromagnetic form factor $F_{\pi}(Q^2)$
appearing in the amplitude
$\gamma^*  \pi^{+(-)} \rightarrow \pi^{+(-)}$,
takes the form of an expansion
\begin{equation}
  F_{\pi}(Q^2)=
   \frac{\alpha_S(\muR)}{4 \pi}
  F^{(1)}_{\pi}(Q^2) 
  + \frac{\alpha_S^2(\muR)}{(4 \pi)^2}
  F^{(2)}_{\pi}(Q^2,\muR) 
  + \cdots
\, .
\end{equation}

\begin{figure}
   \centerline{\includegraphics[width=7cm]{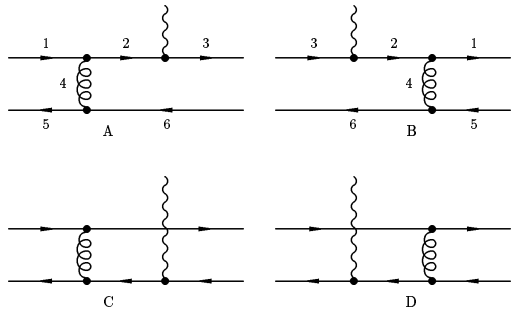}}
\caption{Lowest-order Feynman diagrams contributing to
  $\gamma^* (q \bar{q}) \rightarrow (q \overline{q})$.}
\label{f:LO}
\end{figure}

\begin{figure}
 \centerline{\includegraphics[width=8cm]{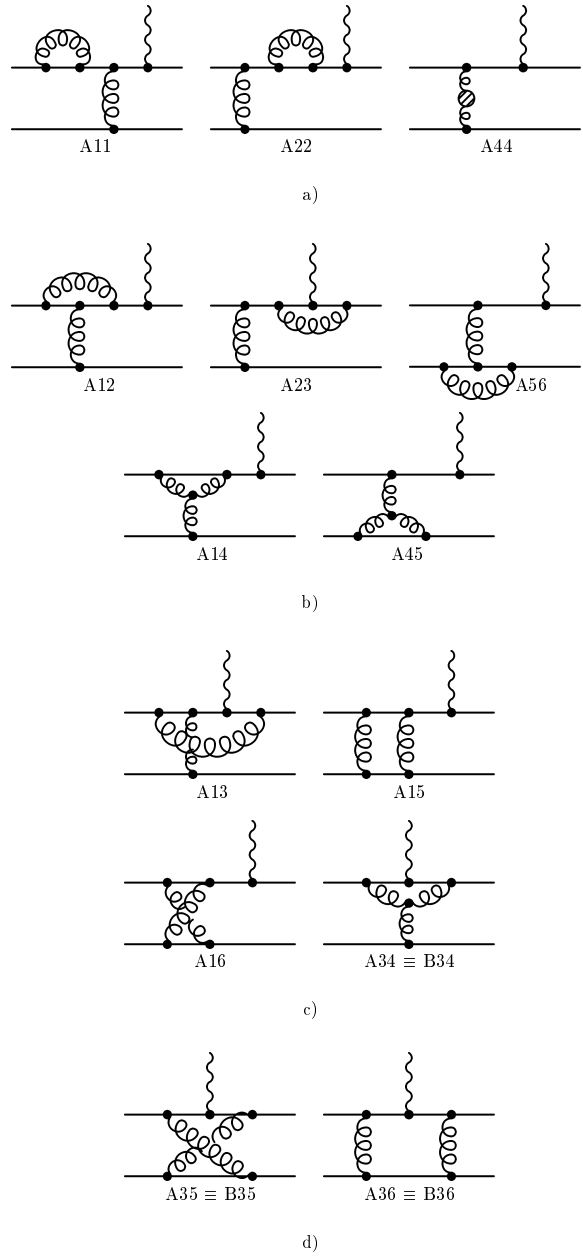}}
 \caption{Distinct one-loop Feynman diagrams contributing to the
  $\gamma^* (q \bar{q}) \rightarrow (q \overline{q})$ amplitude.}
\label{f:NLO}
\end{figure}

There are
4  diagrams that contribute to the amplitude 
$\gamma^* (q_1 \bar{q_2}) \rightarrow (q_1 \bar{q_2})$ 
at LO (see Fig. \ref{f:LO}),
and 62 one-loop diagrams at NLO
   \cite{FieldGOC81,DittesR81,Sarmadi82,KhalmuradovR85,BraatenT87,%
     KadantsevaMR86,MNP99}
(see Fig. \ref{f:NLO}).

\begin{figure}
\centerline{\includegraphics[width=9cm]{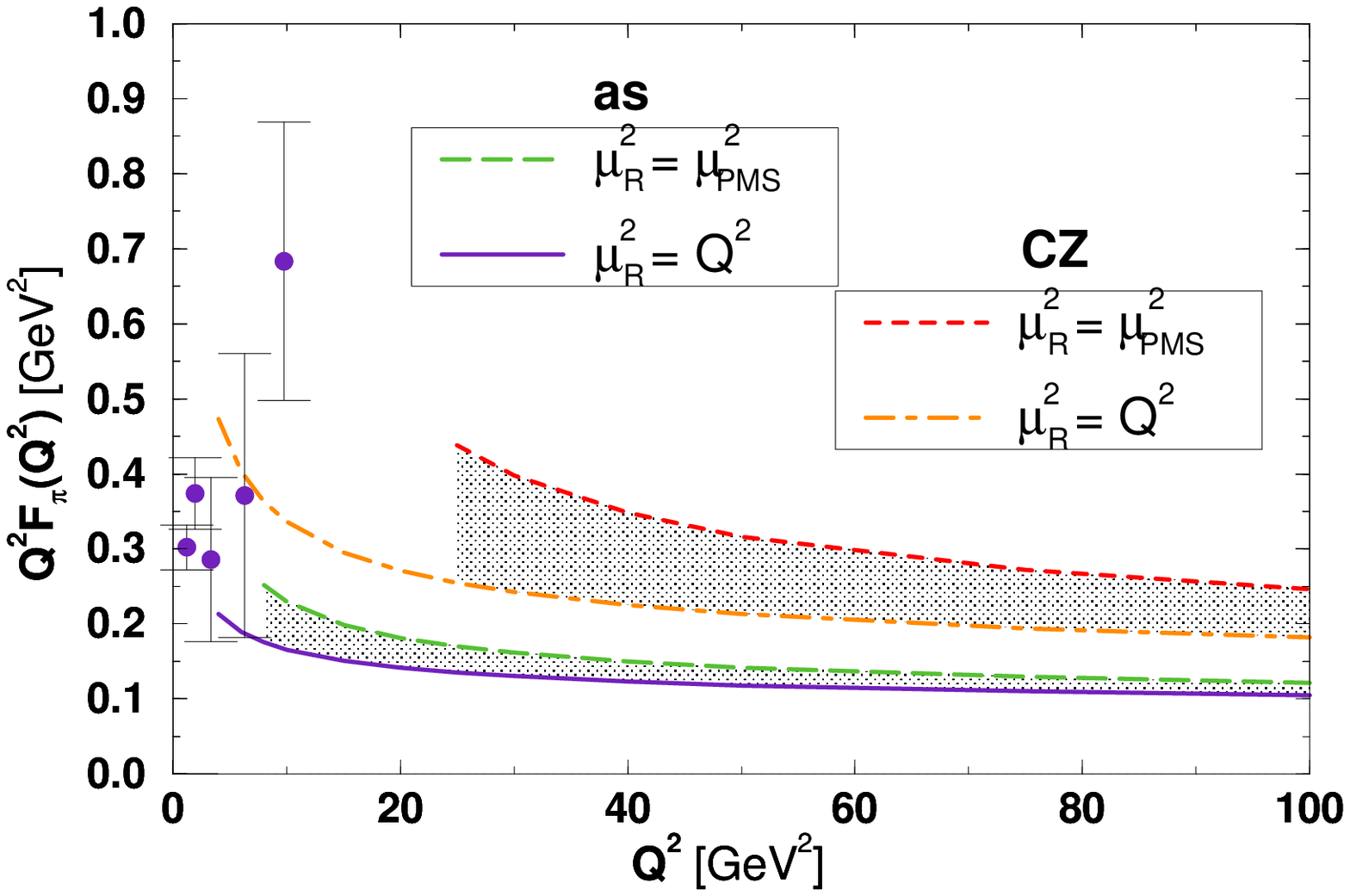}}
\caption{NLO prediction for $F_{\pi}(Q^2)$. The shaded area denotes
the range of the total NLO prediction and offers the way to asses
the theoretical uncertainty.}
\label{f:num2}
\end{figure}

Numerical predictions for $F_{\pi}(Q^2)$ are 
displayed in Fig. \ref{f:num2}.
We comment the asymptotic DA results.
For
$\muR=Q^2$  NLO corrections are rather large: 
the ratio (NLO correction/LO prediction) is 
  $> 30 (50)\%$ until $Q^2>500 (10)$ GeV$^2$ is reached! 
On the other hand, the BLM scale
$\muR=(\mu_{BLM}^2)^{as}\approx Q^2/106$ is
   very small and hence
    $\alpha_S$ is large.
The $\alpha_V$ scheme offers the possible way out.
In this scheme the scale amounts to
$\muR=(\mu_{V}^2)^{as}\approx Q^2/20$ 
   ($\alpha_S<0.5$ and NLO corrections $< 27\%$ for $Q^2>20$ GeV$^2$)
\cite{MNP99a}.

\subsection{
 Pion pair production}
\label{ssec:MM}

Finally, the amplitude of the process
${\gamma} {\gamma} { \rightarrow} {\pi^{+}} {\pi^{-}}$
takes the form
\begin{equation}
{\cal M}(s,t) = 
 \frac{\alpha_S(\muR)}{4 \pi} {\cal M}^{(1)}(s,t)
+ \frac{\alpha_S^2(\muR)}{(4 \pi)^2} {\cal M}^{(2)}(s,t,\muR)
+ \cdots
\, .
\end{equation}

There are 20 diagrams that contribute at LO order
$\gamma  \gamma \rightarrow (q_1 \bar{q_2}) (q_2 \bar{q_1})$.
At NLO order 454 one-loop diagrams contribute to the
NLO prediction.
The existing result from the literature 
   \cite{Nizic87} covers only the
special case of the equal momenta DA, i.e.,
$\phi(x)=\delta(x-1/2)$. The numerical result is thus
not particularly realistic.
New (general) NLO calculation is in preparation
and for that purpose
convenient general analytical method for evaluation
of one-loop Feynman integrals has been developed
\cite{DuplancicN01etc}.

\section{NNLO prediction for the photon-to-pion 
     transition form factor using conformal symmetry constraints}
\label{sec:NNLO}

Recently, the conformal symmetry constraints were used
to obtain the NNLO prediction for the photon-to-pion
transition form factor
\cite{MelicMP02}.
The crucial ingredients of this approach lie in the fact 
that the
massless PQCD is invariant under conformal
transformations provided that the $\beta$ function vanishes,
and that 
$F_{\pi \gamma^*}$ belongs to a class of two-photon
processes calculable by means of the operator product expansion
(OPE).
One can then make use of
the predictive power of the conformal OPE (COPE),
the DIS results for the nonsinglet coefficient function
of the polarized structure function $g_1$ known to
NNLO order \cite{ZijlstraN94} and
the explicitly calculated $\beta$-proportional NNLO terms 
\cite{BelitskySch98,MNP01}.

Let us first introduce the basic ingredients of the formalism.
For the general case of the pion transition form factor
${\gamma^*} (q_1)
\gamma^{*}  (q_2)
 \rightarrow 
\pi(p)$
one expresses the results in terms of
\begin{equation}
\bar{Q}^2= -\frac{q_1^2+q_2^2}{2}
\qquad \mbox{and} \qquad
\qquad
\omega= \frac{q_1^2-q_2^2}{q_1^2+q_2^2}
\, .
\end{equation}
It is convenient to turn the convolution formula
(\ref{eq:Fpigammaconv}) into the sum over 
conformal moments
\begin{equation}
F_{\pi \gamma^*}(\omega,\bar{Q}^2)=
f_{\pi} \,\sum_{j=0}^{\infty}{}' \; T_j(\omega,\bar{Q}^2,\muF) \,
\langle \pi|  {\cal O}_{jj}(\muF) |0\rangle
\, .
\end{equation}
Here
\begin{equation}
T_j(\omega,\bar{Q}^2,\muF) =
\int_0^1 \; dx \; T_H(\omega,x,\bar{Q}^2,\muF) \; 
\frac{x (1-x)}{2 \sqrt{2 N_c} N_j} \;
C_j^{3/2}(2 x -1)
\end{equation}
is the $j$th conformal moment of the elementary hard-scattering amplitude,
while
\begin{equation}
\phi(x,\muF)=\sum_{j=0}^{\infty}{}' \;
 \frac{x (1-x)}{N_j}  C_j^{3/2}(2 x -1)
\langle \pi|  {\cal O}_{jj}(\muF) |0\rangle
\, ,
\end{equation}
where ${\cal O}(\muF)$ represents composite conformal operator,
and $N_j = (j+1)(j+2)/4(2j+3)$.

For the thorough review of the
conformal transformations and their applications
we refer to \cite{BraunKM03}.
On the quantum level conformal symmetry is broken owing to the
regularization and renormalization of UV divergences:
 coupling constant renormalization resulting in
$\beta$-proportional terms and 
the renormalization of composite operators.
The latter represents the origin of
non-diagonal NLO anomalous dimensions in 
$\overline{\mbox{MS}}$ scheme 
and can be removed by finite renormalization
of the hard-scattering and distribution amplitude,
i.e., by the specific choice of the factorization scheme.

First, we pose the question weather we can
find a factorization scheme in which conformal
symmetry holds true up to $\beta$-proportional terms?
Renormalization group equation for the operators
${\cal O}$, equivalent to the DA evolution equation
(\ref{eq:eveq}),
is given by
\begin{equation}
\mu \frac{d}{d\mu}{\cal O}_{j l} =- \sum_{k=0}^j \gamma_{jk} {\cal O}_{kl}
\, ,
\end{equation}
where the
anomalous dimension matrix,
corresponding to the evolution kernel
(\ref{eq:kernel}),
is given by
\begin{equation}
\gamma_{jk} =
\frac{\alpha_s}{2\pi}\delta_{jk}\gamma_{j}^{(0)} +
\frac{\alpha_s^2}{(2\pi)^2}
\gamma_{jk}^{(1)} + \frac{\alpha_s^3}{( 2 \pi)^3} \gamma_{jk}^{(2)}
+ O(\alpha_s^4)
\end{equation}
  with $\gamma_{j} \equiv \gamma_{jj}$. 
Since the conformal symmetry holds at LO,
the anomalous dimensions are diagonal at LO.
Similarly non-diagonal terms present in the
$\overline{\mbox{MS}}$ scheme beyond LO, originate in breaking of
conformal symmetry due to the renormalization of the composite
operators.
In contrast, the conformal subtraction (CS) scheme defined by
\begin{equation}
{\cal O}^{\rm CS} =\hat{B}^{-1} {\cal O}^{\overline{\rm MS}}, \qquad
B_{jk} = \delta_{jk} +\frac{\alpha_s}{2\pi} B_{jk}^{(1)} + O(\alpha_s^2)
\end{equation}
and
\begin{equation}
\gamma^{\rm CS}_{jk} =\delta_{jk} \gamma_j 
+ \theta(j>k)\frac{\beta}{g} \Delta_{jk}
\end{equation}
preserves the conformal symmetry up to $\beta$-proportional terms.

Second, we ask weather and how can we use the predictive power of conformal
symmetry?
The process of interest belongs to quite a large class of
two-photon processes calculable by means of OPE \cite{MullerRGDH94}. 
DVCS, deeply inelastic lepton--hadron scattering (DIS) and
production of various hadronic final states by photon--photon fusion
belong to this class of processes. Such processes can be described by a
general scattering amplitude given by the time-ordered product of
two-electromagnetic currents sandwiched between the hadronic states. For
a specific process, the generalized Bjorken kinematics at the light-cone
can be reduced to the corresponding kinematics, while the particular
hadron content of the process reflects itself in the non-perturbative
part of the amplitude. Hence, the generalized hard-scattering amplitude
enables us to relate predictions of different two-photon processes on
partonic level.

Conformal OPE (COPE) for two-photon processes
works under the assumption that
conformal symmetry holds (CS scheme and $\beta$=0),
and
the Wilson coefficients are then, up to normalization, fixed
by the ones appearing in DIS structure function $g_1$
(calculated to NNLO order).
Conformal symmetry breaking terms proportional to
$\beta$ function alter COPE result.
One can make use of $\beta$-proportional NNLO terms explicitly calculated
in $\overline{\mbox{MS}}$ scheme 
\cite{BelitskySch98,MNP01}.
We note here that there exists freedom in defining $\beta$-proportional
terms in CS scheme 
and hence we speak of CS scheme, $\overline{\mbox{CS}}$ scheme, $\ldots$
(for detailed explanation see \cite{MelicMP02}).

Finally we list the numerical results for the special case
$\omega=\pm 1$ and $j=0$, i.e., asymptotic DA.
For
$\muR=2 \bar{Q}^2$
\begin{eqnarray}
F_{\pi \gamma}(\bar{Q}^2) & = &
\frac{\sqrt{2} f_{\pi}}{2 \bar{Q}^2}
\left[
1 - \frac{\alpha_s{(2 \bar{Q}^2)}}{\pi}  -
     \left\{ {7.23 \atop 5.14}  \right\} \frac{\alpha_s^2{(2 \bar{Q}^2)}}{\pi^2}
\right. \\ & & \left. 
   + O(\alpha_s^3)
\right] \quad \mbox{in}\quad \left\{ { \mbox{CS} \atop
\overline{\mbox{CS}} } \right.\; \mbox{-scheme}\, ,
\end{eqnarray}
while for the BLM prescription
$\muR=\mu_{\rm BLM}^2$
\begin{equation}
F_{\pi \gamma}(\bar{Q}^2) =
\frac{\sqrt{2} f_{\pi}}{2 \bar{Q}^2}
\left[ 1 - \, \frac{\alpha_s{(\mu_{\rm BLM}^2)}}{\pi}  +
      0.92
     \frac{\alpha_s^2{(\mu_{\rm BLM}^2)}}{\pi^2}
   + O(\alpha_s^3) \right]
\,.
\end{equation}
Here
\begin{equation}
 \mu_{\rm BLM}^2 = 2 Q^2 \left\{ {1/37.43  \atop 1/14.78
}\right\}\quad
\mbox{in} \quad \left\{ { \mbox{CS} \atop \overline{\mbox{CS}} }\right.
\mbox{-scheme} \, .
\end{equation}
One notices that, similarly to the pion electromagnetic form factor
results presented in the preceding section,
for $\muR$ equal to the characteristic scale of the process
the QCD corrections are large
\footnote{Note that for the case of the meson transition form factor
NLO correction represents actually LO QCD correction, while
NNLO correction is NLO QCD correction etc.}, while
for the BLM scale these corrections are smaller but the scale
itself is also rather small leading to large expansion parameter
$\alpha_S$. The $\alpha_V$ scheme could as in the case of the 
pion electromagnetic form factor offer the way out and physically 
better motivated description of the transition form factor.

We mention, that,  
as already noticed in \cite{DiehlKV01} 
and shown in \cite{MelicMP02}, the significance
of higher-conformal moments decreases with $|\omega|$ 
and that
with decreasing $|\omega|$ the difference between 
various schemes also decreases.
Hence, small $|\omega|$ region is suitable for 
a novel test of PQCD.

\section{Conclusions}
\label{sec:concl}

Although the higher-order QCD corrections are important,
only few exclusive processes have been
 explicitly calculated to NLO order.
The inclusion of higher-order corrections
stabilizes the dependence on renormalization scale.
Still, the usual choice $\muR=[\mbox{characteristic
scale of the process}]$ leads to large corrections.
Other choices of scales 
(BLM, $\alpha_V$ scheme) are preferable and more physical.
More effort in calculating higher-order corrections
are needed and some tools applicable to
the kinematic region of interest are underway.
Furthermore, for some processes 
(example: NNLO calculation of photon-to-pion transition form factor), 
one can make use of the predictive
power of conformal symmetry to avoid cumbersome higher-order calculations.

\section*{Acknowledgments}

I would like to take the opportunity to thank 
P. Kroll, B. Meli\'{c}, D. M\"{u}ller and B. Ni\v{z}i\'{c}
for fruitful collaborations in challenging higher order
calculations, my collaborators A. P. Bakulev, N. G. Stefanis and W. Schroers 
for interesting extensions of these investigations, 
as well as, H. W. Huang, R. Jakob, M. Sch\"{u}rmann
 and W. Schweiger for collaborating on
stimulating and demanding LO calculations.
This work was supported by the Ministry of Science and Technology
of the Republic of Croatia under Contract No. 0098002.

\end{document}